# To What Extent Is Stress Testing of Android TV Applications Automated in Industrial Environments?

Bo Jiang, Peng Chen, W.K. Chan†, and Xinchao Zhang

*Abstract*－An Android-based smart Television (TV) must reliably run its applications in an embedded program environment under diverse hardware resource conditions. Owing to the diverse hardware components used to build numerous TV models, TV simulators are usually not high enough in fidelity to simulate various TV models, and thus are only regarded as unreliable alternatives when stress testing such applications. Therefore, even though stress testing on real TV sets is tedious, it is the de facto approach to ensure the reliability of these applications in the industry. In this paper, we study to what extent stress testing of smart TV applications can be fully automated in the industrial environments. To the best of our knowledge, no previous work has addressed this important question. We summarize the findings collected from 10 industrial test engineers to have tested 20 such TV applications in a real production environment. Our study shows that the industry required test automation supports on high-level GUI object controls and status checking, setup of resource conditions and the interplay between the two. With such supports, 87% of the industrial test specifications of one TV model can be fully automated and 71.4% of them were found to be fully reusable to test a subsequent TV model with major upgrades of hardware, operating system and application. It represents a significant improvement with margins of 28% and 38%, respectively, compared to stress testing without such supports.

*Keywords—Stress Testing, Android, TV, Reliability, Automation, Test Case Creation, Software Reuse*

**Acronyms**

| | |
|---|---|
| TV | Television |
| CPU | Central Processing Unit |
| MIC | Microphone |
| ADB | Android Debug Bridge |
| D-pad | Directional Pad |
| GUI | Graphical User Interface |
| API | Application Programing Interface |
| OS | Operating System |
| USB | Universal Serial Bus |
| TTS | Text To Speech |
| SAPI | Speech Application Programing Interface |
| ANR | Application Not Responding |
| TAST | Testing of Android-based Smart TVs |
| ANOVA | Analysis of Variance |

- Bo Jiang is with School of Computer Science and Engineering, Beihang University, China. E-mail: jiangbo@buaa.edu.cn.
- Peng Chen is with School of Computer Science and Engineering, Beihang University, China. E-mail: pc@buaa.edu.cn.
- W.K. Chan is with Department of Computer Science, City University of Hong Kong, Tat Chee Avenue, Hong Kong. E-mail: wkchan@cityu.edu.hk.
- Xinchao Zhang is with Institute of Reliability, Si Chuan Chang Hong Electric Co., Mianyang, China. E-mail: zhangxinchao@changhong.com

† Correspondence author.

*Manuscript received (insert date of submission if desired). Please note that all acknowledgments should be placed at the end of the paper, before the bibliography.*

## I. Introduction

Smart televisions (Smart TVs) [11] are widely-used embedded systems [10][13], and a major class of such embedded system is *Android-Based Smart TVs* [29]. Different models even for the same series of TV use diverse types of hardware components. Each model in this class includes much standard TV functionality, such as channel controls and preferences, as well as non-standard applications such as online game clients, web browsers, multimedia players, photo albums, in the form of Android applications, and executes them concurrently. At any one time, users may turn on zero or more non-standard applications and keep these applications executing while watching TV, or the other way round. As we are going to present in Section II.A, stress testing of TV applications can be quite different from the testing of applications of Android smart phones. Moreover, a high degree of test automation is in high demand in industrial environments.

We thus ask a couple of research questions. (i) What are the areas that test engineers consider most important, which require automation when stress testing TV applications? (ii) To what extent do semi-automated test cases become fully automated for the purpose of stress testing?

To the best of our knowledge, no previous work has studied these two important questions. To answer these two questions, in this paper, we report our 7 months case study on improving the degree of automation in the stress testing of TV applications in a major TV vendor (Changhong [27]) in China.

Specifically, we worked together with ten (10) test engineers in testing 20 real-world applications in a real-world smart TV model manufactured by the same vendor for 7 months (Mar−Sep 2013) using real TV sets. These test engineers have 2 to 5 years of TV testing experiences with an average of 3 years. They have experiences on both digital TV testing and smart TV testing. Each test engineer was assigned to test two TV applications for a TV model. Because the TV model is not the first model in the TV series, each application was associated with a set of test scripts. Our methodology was to firstly observe how test engineers used their original Android testing toolkit (denoted by *MT*, which was an upgraded version of MobileTest [7] for Android) to create and maintain their automated test scripts based on the corresponding test case specifications (or *test specification* for short). Following a typical recommended practice of code refactoring, when we observed that they repeatedly wrote similar code fragments, we asked questions on what they wanted to automate further to save their repetitive efforts. Moreover, on observing them aborting the automation of a test specification, we also asked questions on why that particular test specification was failed to be automated and what automation features to be available in order to make them willing to successfully write automated test scripts. We then added each identified feature to the same testing tool as new APIs if the feature either is a piece of code produced via "method extraction" in code refactoring or controls the available hardware resources in the program execution of a test script. We continuously enhanced and released the testing tool with added features, and the test engineers continuously used the latest releases of the testing tool to create and maintain their test cases. We note that, as new APIs were available to test engineers, their subsequent test scripts might incorporate additional coding patterns that triggered the discovery of new APIs to be incorporated into newer versions of the testing tool. We continued this process of tool enhancement until all 10 engineers found further automation unable to assist them in testing their assigned TV applications meaningfully (in their work environments). We then requested each test engineer to review the possible uses of the original version of the testing tool (i.e., MT) to create test script(s) that were originally chosen not to be automated by the test engineers.

Finally, we measured the effects of the resultant test suites of these 20 applications produced through the latest version of the testing tool against the effects of the test suites using the original version of the testing tool (MT). For the ease of our reference, we refer the latest version of the testing tool to as **TAST**. We also note that the TV model tested via the latest version of the testing tool in the case study was being sold in the Mainland China market at the time of writing this article.

The results of the case study show the importance of providing a higher level of abstraction with resource controls in the stress testing of TV-based applications on real TV sets. First, test engineers in the case study examined 1563 test specifications in total. Originally, with the assistance of MT, they fully automate 915 test specifications. On the other hand, with the assistance of TAST, they fully automate 1347 test specifications. The difference is 432, which represents 50% *more* test cases. Moreover, using TAST, test engineers fully automated 75.0–96.3% (with a mean of 86.8%) test specifications of these 20 applications. This degree of automation was significantly higher than that using MT, which only achieved 50.0–71.3% (with a mean of 58.7%).

Second, when applying the resultant test scripts to conduct a session of stress test on a newer version of the same TV model series, we found that 55.0–88.0% (with a mean of 71.4%) of TAST test scripts can be completely reused. The amount of complete reuse is significantly higher than that achieved by MT, where only 31.0–49.0% (with a mean of 40.7%) of the MT test scripts can be reused. Furthermore, out of these reusable test cases, 59–63% of them do not cover any change in code and 37–41% of them cover some changes in code and can still work correctly. Note that the newer TV set has used newer operating system and hardware, making the program environments not the same as the ones perceived by test engineers when they wrote test scripts based on an older TV set.

Third, being able to produce more automated test scripts means being able to produce more scenarios for stress testing. Stress testing required the execution of many test scripts extensively. We also found that test scripts provided via TAST had exposed previously unknown bugs during the above-mentioned iterative process. Specifically, we found that the TAST test scripts exposed real faults from each application with a failure rate of 1.9−4.1% (with a mean of 3.2%). We emphasized that the TV model analyzed in the case study was not the first model in the TV series to use the same test specification for both stress testing and functional testing. This result showed that the new automated test scripts exposed unknown bugs that were not exposable before the arrival of TAST.

Last, but not the least, the case study confirmed and validated the feasibility of the above methodology to produce an effective testing framework.

The main contribution of this paper is twofold: (i) This paper presents the *first* work on studying to what extent stress testing of TV applications can be automated. (ii) It reports the *first* large-scale industrial case study on the stress testing of Android applications in the industrial environment.

We organize the rest of this paper as follows: Section II presents the motivation, including a motivating example, of our work. Section III describes the design of the resultant testing tool. Section IV and Section V present a case study that validates the methodology and investigates the research questions that motivate this work. Section VI reviews the closely related work. Finally, Section VII concludes this paper.

## II. MOTIVATION

### A. Inadequacy of Testing Infrastructure

Testing smart TV applications is significantly different from testing smart phone applications. In this section, we present four areas of differences that we observed from conducting our case study to motivate this work further.

First, TVs are typically designed to be viewed 10-feet away, and TV applications should provide the so-called "10-foot user experience" ([18], page 42), where user interface (UI) objects such as scroll bars may be too small to precisely control by users. As such, screen-based or touch-based inputs are seldom used; and the UI interaction of TV applications is mainly controlled by D-pads (Directional pads), which usually contains the left key, right key, up key, down key, select key, as well as receiving voice commands. Hence, traditional Android application testing tools, which issue the UI touch event to drive the device, are often invalid to test TV applications. The need of voice-based input demands for automatic voice input testing.

Second, traditional TVs are mature and reliable through many decades of development. End users of smart TVs include people of all ages and all educational backgrounds, and they all require the TV (including its TV applications) to be reliable. Stress testing of TV applications should be much more comprehensive (e.g., including all sorts of dialect variations and voice tones of the same speaking language when testing against the voice input channel) than stress testing of typical Android applications.

Third, unlike a phone simulator sufficient to test a typical Android phone application, TV simulators are seldom high in fidelity, making them as inferior and unreliable alternative to real TV sets when stress testing TV applications. Therefore, even though stress testing on real TV sets is tedious, it is the de facto approach to ensuring these applications in the industry. Nonetheless, controlling a real TV set is much more tedious than controlling a simulator.

In summary, a testing tool for smart TV testing differs from the testing tool for smart phone testing in three aspects: first, it calls for D-pad input support and voice input support. Second, stress-testing support is crucial to ensure the TV is of high reliability for selling as consumer electronics. Third, smart TV lacks a high fidelity simulator, which demands the testing tool to support testing on real TV sets.

*B. Motivating Example*

We use an example to illustrate the challenges in the stress testing of an application embedded in a smart TV.

The scenario is as follows: Suppose that a Smart TV system has the following two applications: a media player (denoted by $A_1$) and a voice assistant application (denoted by $A_2$), which translates an audio stream into commands. Suppose further that a user is watching a local video using the media player $A_1$. Ideally, upon the user pronouncing a word (e.g., T-V-B) to the smart TV, the application $A_2$ should receive and analyze the sound followed by switching the TV screen to some other applications (that shows the targeted particular TV channel).

Conducting a session of stress testing on $A_2$ can be challenging and tedious. To make our presentation concise, we refer to a test engineer in the following testing scenario as *U*. First, *U* manually navigates on the application launch panel of the TV to activate $A_1$ to watch a local video clip. *U* then sets up a resource-constrained Smart TV environment to prepare for the testing of $A_2$ running the TV as follows: To reduce the CPU and memory resources available to $A_2$, *U* invokes some computationally intensive applications and invokes numerous other applications, respectively. The reason is that each Android application tends to behave like a small Linux process (e.g., occupying 20MB memory only), and a Smart TV may easily have two orders of magnitude more memory than the need of the former (e.g., 8GB). A Smart TV is usually designed to launch a limited number of applications. As such, *U* requires installing many other applications in the Smart TV under test to occupy internal memory, so that the memory left for execution of the targeted application can be "small enough". Installing applications on a Smart TV involves downloading these applications from Android *app markets* (over the internet) with a process of user confirmation each. Each installation however takes some time (e.g., 1 minute). Installing hundreds of applications in the Smart TV is thus a tedious process. After setting up the environment, *U* spells the required word according to a test specification, and visually observes whether the TV screen switches to the required TV channel correctly and in time. The main problems are that the steps involved are semi-automated, tedious, inaccurate, and non-representative to the operating situations.

```
1    # -*- coding: utf-8 -*-
     /*Import the required TAST packages*/
2    from com.android.tast import TAST
3    from com.android. tast import MTDevice
4    from com.android. tast.easy import By
5    from com.hierarchyviewer import HierarchyViewer
     /*Connecting to the device under test*/
6    device = MobileTest.waitForConnection()
     /*Start the media player application */
7    device.startActivity("com.mediaplayer.MediaPlayerActivity")
8    MobileTest.sleep(5)
9    /*Start a stress testing agent to consume target amounts of
     resources*/
10   device.consumeMemory(90)
11   device.consumeCPU(90)
12   device.consumeNetwork(90)
13   MobileTest.sleep(10)
     /*Check and print resource consumed to console*/
14   memoryInfo = device. getMemoryUsage()
15   print memoryInfo
16   netInfo = device.getNetworkUsage()
17   print netInfo
18   cpuInfo = device.getCPUUsage()
19   print cpuInfo
     /*Specify a command for the voice assistant app */
20   voices = [u'Channel 4', u'Star Trek', u'Weather of Beijing',
     u'music of Taylor Swift', u'Search Google']
21   oracle=[ u'com.atv.activity.AtvMainActivity',
             u'com.mediaplayer.MediaPlayerActivity',
             u'com.changhong.app.WeatherActivity',
             u'com.changhong.onlineMusic',
             u'com.android.browser.BrowserActivity']
22   for index, str in enumerate(voices):
23   device.sendVoiceCmd(str)
24   activity = device. getFocusedWindow ()
         /*Verify whether the correct application invoked */
25   device.assertion(activity.getName() == oracle[index])
     /*Tear down the test script by going back to the Main
     Window */
26   device.press ("KEYCODE_BACK")
27   /*Release the consumed resources */
28   device.StopConsumeMemory()
29   device.StopConsumeNetwork()
30   device.StopConsumeCPU()
31   #MTRecorder@end
```

Figure 1. A fully automated TAST test script for the stress-testing scenario in the motivating example

Our goal is to understand and testify to what extent automation can be feasible. Following the methodology presented in the last section, TAST simplified the testing steps involved by allowing *U* to develop fully automated test scripts. Figure 1 shows a fully automated TAST test script that serves the same purpose as the above tedious testing process: After importing some required TAST libraries (lines 2 to 5), the script connects the TV under test (line 6) to instantiate a *device* object. It starts an instance of $A_1$ (line 7) and controls the execution environment of the *device* object by consuming 90% of CPU, memory, and network resources (lines 9 to 12). It then shows the resource consumption status (lines 14 to 19), and declares a sequence of voice commands and the test oracle for the command, respectively (lines 20 and 21). It sends each voice command sequentially to the *device* object, and verifies whether the corresponding applications have started successfully according to the above-defined test oracle (lines 22 to 25). Finally, the script releases the TV resources deliberately held by TAST (lines 28 to 30). In TAST, the resource control is abstracted as properties of the *device* object. From the execution of line 10 to line 30, the CPU utilization level of the Smart TV is actively sustained at 90% (i.e., only 10% of the CPU resources are available to the execution from line 10 to line 30). Moreover, the communication between the Smart TV and TAST is completely abstracted away from the test script, and feedbacks from the Smart TV (e.g., memory usage or the active GUI control objects) are abstracted as data objects (e.g., *memoryInfo*) in the test script. As such, the test script can use the property of such an object to control the execution of the test script.

There are two notes on this motivating example that deserve explanation. First, the motivating example mainly focuses on presenting why our TAST framework can automate some manual testing effort. We will discuss the reusability aspect of our TAST platform in the RQ2 of our case study. Second, the use of hard-code value is mainly for improving the readability of the test script to ease understanding. In fact, we have also provided a data-driven interface (API) for TAST, where the test scripts can read input data from files or databases to enable data-driven testing.

### III. THE RESULTANT TESTING TOOL: TAST

In Section 1, we have presented the methodology to identify a feature needed for automation. In this section, we present the resultant tool produced via the methodology. We firstly present the overall design, and then enumerate each kind of features that we have identified.

#### A. Overview of the TAST Architecture

In this section, we present the overall architecture of TAST. As depicted in Figure 2, TAST consists of four layers in additional to an agent service embedded in the device.

At the bottom layer, TAST communicates with a Smart TV via a set of TAST-TV interfaces: the ADB interface [24], a Socket-based voice interface, and a Socket-based Agent interface. The former two interfaces provide the input/output channels to mimic user interactions with the application under test (AUT) in the TV, and the third interface enables TAST to *control* the available resource to the execution environment of the AUT. These three TAST-TV interfaces are connected to three controllers (ADB Controller, Voice Controller, and Agent Controller, respectively) of TAST as shown in Figure 2.

The first TAST-TV interface known as the ADB (abbreviated for Android Debug Bridge) [24] interface is a standard interface common to all Android-based systems. Through this interface, the ADB controller in the next layer above controls a built-in testing component to issue user events to the AUT and capture screenshots of the AUT running in the Smart TV.

We note that most testing tools for Android applications share the above ADB interface. Our MT testing platform (i.e., MobileTest) has no exception. Like TAST, MobileTest was also built on the services provided by

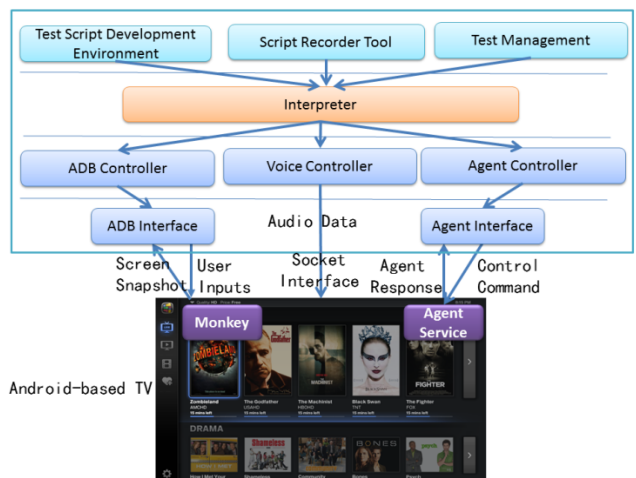

Monkey.

Figure 2. Architecture of our testing platform TAST

Owing to the need of voice control, simply using ADB is insufficient to provide the inputs to some TV applications adequately. Therefore, in response to the testing requirement, we added a new TAST-TV interface (the Socket-based voice interface) to the testing tool. The purpose of this interface is to send voice control commands to interact with the voice assistant application (e.g., Ciri [28]) in a smart TV.

The test engineers expressed that stress testing of TV applications requires setting up the resource available for the applications to use. Therefore, we introduced a third TAST-TV interface (i.e., the Socket-based *Agent* interface) to the tool, which is to communicate with an *agent service* of TAST running on the Smart TV to control and monitor the resource utilization levels of various resources.

The layer above the TAST-TV interface layer is the controller layer, where we have the ADB-controller, voice-controller, and the Agent controller. The ADB controller encapsulates the ADB interface, the voice-controller encapsulates the socket-based voice interface, and the agent controller encapsulates the agent interface. These three controllers then each provide a set of interface to support test script execution controlled by the interpreter in the layer above. This controller layer essentially encapsulates the interactions with the Android device such that the interpreter above can focus on the interpreted execution of test scripts per se.

The execution of each test script is handled by a test script interpreter, which forms the second top layer as shown in Figure 2. The interpreter handles instructions sequentially along the execution trace of each test script. For every such instruction, it translates the instruction into lower-level tasks, interacts with the Smart TV via the above set of TAST-TV interfaces to carry out all these tasks.

The top layer of TAST shown in Figure 2 is for test engineers to develop test scripts such as the test script shown in Figure 1. The test management component is an abstraction for organizing and executing test scripts.

The test script development environment and the script recorder tool are also in the top layer. The former provides an integrated test development environment for the test engineers to develop test scripts while the latter is a classic test-script recorder tool by logging user interactions.

In the next section, we will present the new APIs identified through the methodology presented in Section 1.

### B. Events and Execution Trace Model

We have also highlighted in Figure 1 that TAST provides methods (e.g., `consumeMemory()`) that are specifically related to the TAST-TV interfaces (see Section 3) in the form of its Application Programming Interfaces (API). As such, during the test script execution, the interpreter invokes a sequence of such API methods. We refer to every request to invoke such an API method as an *event*, and hence an execution trace is viewed as a sequence of such events. The interpreter executes along an execution trace sequentially.

The interpreter catches all unhandled exceptions thrown by a test script during the execution of the test script. It aims at rendering a layer of fault tolerance to the test environment of the AUT to improve the reliability and usability of TAST. Suppose that the unhandled exception occurrence is from the ADB Controller (occurred when the ADB connection is unexpectedly inactive). The interpreter will invoke the `resetADB()` method that resets the ADB connection followed by re-sending the event that generates the unhandled exception within a predefined number of trial attempts. Similarly, suppose that an unhandled exception occurrence is from the Agent Controller. The interpreter invokes another method, which is `resetAgent()`, to restart the agent service residing in the smart TV, and looks back the execution trace to identify latest resource consumption events and reissue these events to the agent service so that the agent service can re-establish the required environmental resources setting. (We note that the number of trial attempts is a configuration parameter defined by tool users.) For other exception occurrences, the interpreter forwards the exception occurrences to its own program environment. The interpreter also throws an exception if any assertion statement (e.g., line 25 in Figure 1) is violated.

The test management component catches all exceptions thrown from the interpreter followed by marking the test script execution as *failed*. Otherwise, if the test script is executed without any exception, it marks the test script as *passed*.

TAST defines three parallel TAST-TV interfaces. Correspondingly, there are three *primitive* types of events, one for each TAST-TV interface. In the next three subsections, we present these three types of events in turn.

### C. Events for the ADB Interface

The ADB interface links the ADB controller in TAST and a *monkey* [31] resided in an Android-based TV. Like MT, TAST has been, by default, configured to operate with the monkey provided by Android SDK. This monkey listens to a default network port on the device to accept text-based commands. It executes one text-based command and returns a text-based output, before executing the next text-based command received. For instance, the monkey accepts a variety of commands such as drag, mouse click, key press, install/remove package. The state information of a particular control object of an Android application can be obtained by sending the DUMP command or the `DUMPQ` command (a lightweight DUMP implementation) [31].

However, unlike MT, in TAST, for each text-based command that the monkey accepts as an input, the ADB controller wraps the command as a corresponding method in the TAST API with the sequence of parameters that exactly matches the types and the sequence of the parameters used by this command. For instance, the commands to represent making a connection to the TV, pressing a specific key, and clicking a specific mouse button are modeled as `waitForConnection()`, `press()`, and `mouse()`, respectively. This allows test scripts to communicate with the monkey in a higher level of abstraction.

We note that each method directly mapped from each command of a monkey is quite primitive (and low level), which is exactly the type of APIs provided by MT. Although test engineers may use such wrapped methods to control a monkey in TAST test scripts in full strengths, yet instructing the monkey to complete a testing-oriented task via such a set of primitive methods are reported by test engineers to be tedious, lengthy, and non-productive, which is illustrated below.

Let us consider a scenario. Suppose that a test script needs to acquire the identity of a particular GUI window, which is the current focusing window among all Android applications in the Smart TV. In the body of the test script, the test engineers need to write instructions to invoke an API method corresponding to the `GET_FOCUS` command of the monkey to retrieve a string of texts, in which the string lists out the *hash code* of the currently focused window. The test script should then extract this hash code. Next, the test script should invoke another API method corresponding to the `LIST` command of the monkey to the view server of the Android operating system to retrieve a list of currently active windows, which is again in the text string format.

The test script then parses the text string to identify the window identity fragments, and compares the text of each window identity in turn to the above hash code to identify the required window identity. In other words, to complete this task, it is not merely a few method invocations of such methods; rather, this task involves the development of program code to link up these method invocations.

Table I. Questionnaire to Test Engineers

| No. | Questions |
|---|---|
| 1 | Are there similar code fragments that you repeatedly write in different test scripts? |
| 2 | What combinations of the testing APIs you use frequently them together? |
| 3 | List some testing APIs you feel cumbersome to use? |
| 4 | When writing test scripts, which part of the code often costs you most of the time? |
| 5 | What are the test specifications that you often give up to automate? |
| 6 | What are the test specifications that you feel hard to automate? |
| 7 | What automation features you desire most in a future version of testing framework? |
| 8 | For semi-automated test scripts, please list examples of the manual efforts required. |
| 9 | For manual test scripts, which part of the execution process requires most manual efforts. |

As mentioned in the introduction section, to determine the automation features to support by our TAST tool, we asked a set of prepared questions to the test engineers following a typical recommended practice of code refactoring. We summarize the questionnaire in Table I.

Based on the analysis on the collected answers of the questionnaires and some inspections on existing test scripts, we paired with the test engineers to identify a set of real world, useful code templates. Each code template is finally wrapped as a method in the TAST API. For instance, the functions in the above scenario have finally been wrapped as one single method `getFocusedWindow()`, which not only obtains the window identity, but also creates (and returns) an object instance in the test script execution state that represents a corresponding control object of the "remote" AUT. The testing tool also automatically maintains the relations between this control object and its associating application and exposes the relationship to the execution state of the test script. Also, to get the GUI control object with a specific control identity, there is a method entitled `getControlItemOp()`, which sends a `DUMPQ` command to the monkey to query the view server to obtain a tree of GUI controls. It then searches the tree to locate the required control object identity, and returns an object that model the matched control object. Hence, a test script can manipulate the control object or its associating application object using much fewer and simpler code. Similarly, TAST provides methods to query the GUI states at all levels: the whole system level, single application level, single window level, single control level, and single property level. Invoking a lower level of query consumes less time before returning the result. Developers may invoke selective methods to implement their own test oracle strategies in their test scripts. TAST currently provides 36 such high-level GUI-object methods in total.

The above scenario also illustrates that the text returned by a command (e.g., `DUMPQ`) may contain much information and there are programming efforts to extract the required pieces of texts from the text. TAST provides several API methods, each finer in granularity than what an underlying command provides. For instance, `getText()`, `isFocused()`, and `isVisible()` are API methods to query the corresponding properties of the control objects inputted as their parameters. In total, TAST currently provides 20 such high-level GUI-attribute methods.

We have presented how TAST handles the events (API methods) that each instructs TAST to query the current states of some control objects and pass back these states to the interpreter. Sometimes, a test script requires methods that each waits for the occurrence of a particular GUI state to appear. For instance, if a window is still invisible before a text input event to a control is received, the event will be lost, making the testing invalid.

TAST provides methods (e.g., `WaitingUIState()`) for each level of GUI object (including application, window, and control) to wait for the occurrences of the corresponding GUI states. Specifically, if the interpreter generates such an event, the ADB controller checks whether the GUI object has changed to the expected state as specified in the method parameter every now and then. This period, say 3 seconds, is configurable in TAST. A match will resume the execution of the interpreter. Otherwise, if a timeout event raised by the interpreter occurs, the interpreter will raise an unhandled exception. With such a set of methods, a test script can use non-sleeping statements to synchronize their testing commands with the corresponding GUI controls.

We note that if a sleep command is issued, the parameter needed for one TV model may require fine-tuning, but then the parameter may not be applicable to other TV models (e.g., using processors with more cores) or operating system versions and other suites of applications to be installed in the TV. They may make the test scripts significantly less non-reusable.

In summary, we found that to support test scripts that manipulate applications through their GUIs, there require supports to identify GUI elements at different levels of details. We have discovered 36 API methods to query GUI states from the whole TV set level to individual property level, which hides the internal navigation of the GUI object structures obtained from ADB from test engineers when using such APIs. This allows a more flexible coding support on checking and setting the GUI objects. We have also discovered 20 API methods each to extract partial information from a lower level of command. We found that synchronization between a state of a test script and a particular state of a particular GUI object is mandatory, which has been frequently missed in the test scripts generated by existing stress testing tool like MT. We shared this finding with test engineers, and they agreed that the finding was consistent to their first-hand experiences.

## D. Events for the Agent Interface

The agent interface bridges between the agent controller and an agent service on the Android system deployed by the test engineers. When the interpreter generates an event for the agent interface, the interpreter will pass the event to the agent controller of TAST. There are two types of events, one for monitoring the resource utilization level of the Android system and another for active setup and maintenance of these resource utilization levels. Any failure in processing such an event (e.g., unable to attain a required memory usage level specified by the event) will trigger an exception by the agent controller.

*Profiling events:* Our industrial collaborator specifies that there are five typical kinds of resource information on an Android system that they need to know in order to test Android applications in their industrial environment: memory usage statistics, CPU usage statistics, network usage statistics, USB storage usage statistics, and operating system (OS) information. We want to ensure our tool to acquire the same information as their own profiling procedures. Therefore, we asked the test engineers to provide their procedures to acquire such information.

```
   /*Function for consuming the CPU usage */
1  Input: the percentage of the CPU to consume.
2  Output: The CPU usage are consumed as required.
3  void consumeCPU(int percentage){
4  /*get the current CPU Usage */
5     float currentPercent = getCPUUsage();
6     float toConsume = percentage – currentPercent;
7     if( toConsume > 0){ /* need to consume resouces */
8        int num = numofCores();  /* get the number of cores */
9        /* start a thread for each core */
10       for(int i=0; i<num; i++)
11          new CPUServiceThread(toConsume).start();
12    }/*if*/
13 }/* end of ConsumeCPU */
14 private class CPUServiceThread extends Thread {
15 private int tPercent;
16 private boolean stop;
17 public CPUServiceThread(int percent) {
18       stop = false;
19       tPercent = percent;
20    }
21  public void run() {
22     long time;
23     /* if not yet stopped by the script*/
24     while (!stop) {
25          time = System.currentTimeMillis();
26          while (System.currentTimeMillis() - time < 10);
27          try {
28             Thread.sleep(10 * (100 - tPercent) / tPercent);
29           } catch (InterruptedException IException) {}
30        }/* while*/
31    }/* run*/
32   public synchronized void stopThread() {
33        stop = true;
34 }
```

Figure 3 Algorithm consumeCPU() for controlling CPU usage

We model their profiling procedures as methods in an extensible class hierarchy of TAST. Specifically, TAST wraps the Linux commands "`cat /proc/meminfo`" and "`top`" as the classes for profiling the memory and CPU usages as `getMemoryUsage()` and `getCPUUsage()`, respectively. To get the network usage information such as the uplink and the downlink network speeds, TAST uses the Android API **android.net.TrafficStats** class to get the data sent and received per second, and calculate the current network speed (by adding up these two values) accordingly in the method `getNetworkUsage()`. The method `getUSBUsage()` invokes the methods of the Android API **android.os.Environment** and **android.os.StatFs** classes to get the path of a USB device as well as its total space and available space. Lastly, the method `getOSInfo()` directly wraps the Android API **android.os.build** class. All these events are sent through the agent interface to the agent service, which in turn invokes these Linux commands or Android APIs to obtain the corresponding outputs, and sends the outputs back to the agent controller. (We note that the agent service in the TV set is also an Android service.)

*Controlling events:* It is essential to set up and *maintain* the amount of resources consumed to the level as specified by an instruction in a test script. We are going to describe the strategy used by TAST for this purpose to control memory, CPU, network, and USB usages.

```
   /*Function for consuming the network bandwidth*/
1  Input: the percentage of the network bandwidth to
   consume.
2  Output: The network bandwidth is consumed as required.
3  void consumeNetwork(int percentage){
4     /*calculate the number of network consumer threads */
5     int num=MAX_NETWORK_THREADS*percentage/100;
8     /* start the threads accessing files on USB storage */
9     for(int i=0; i< num; i++)
10        new NetworkAccessThreads(i).start();
11 }/*end of consumeNetwork*/
12   private class NetworkAccessThreads extends Thread {
13      private int idx;
14      private boolean stop;
15      public NetworkAccessThreads (int aIndex) {
16         stop = false;
17         idx = aIndex;
18      }
19    public void run() {
20      /* if not yet stopped by the script*/
21      while (!stop)
22      /* even numbered thread send data*/
23      if( idx % 2 == 0)
24         Post data to Web server with HTTP client;
25      else /* odd numbered thread receive data*/
26         Get data from Web server with HTTP client;
27    } //run
28    public synchronized void stopThread() {
29        stop = true; } /* stopThread*/
30 } /* end of NetworkAccessThreads*/
```

Figure 4. Algorithm consumeNetwork() for controlling network usage

To keep the memory consumption at a specified level (e.g., 90% of all memory), the agent service actively allocates and de-allocates memory blocks via Linux's native memory management library through the Java Native

Interface so that it bypasses the memory usage restriction imposed by the Android OS on the agent service. A memory usage control is valuable in testing memory-intensive applications such as games. TAST wraps the procedure as the method `consumeMemory()`. For instance, it is critical to test whether an application runs or shuts down correctly even in an execution environment with a small amount of available memory.

A Smart TV typically is equipped with a multi-core processor (CPU). Figure 3 shows the `consumeCPU()` algorithm to control the CPU usage. The algorithm first estimates the total amount of additional CPU loads to be consumed (lines 5 to 6). Then, it starts the same number of threads as the number of CPU cores (lines 8 to 11) For each thread, the algorithm uses a busy loop and the sleep system call to consume a certain percentage of the CPU processing capability and release the consuming CPU capability, respectively (lines 25 to 28). We have found that in our industrial case study, this algorithm can effectively control the CPU usage between 10% and 98% when testing a TV application. This is extremely useful for stress testing a computationally intensive application such as playing or recording a video. In an environment with low CPU availability, these applications should either degrade their quality of services or quit gracefully instead of crash or becoming non-responsive to users.

Figure 4 shows the `consumeNetwork()` algorithm. In essence, to control the network bandwidth usage, the algorithm starts several threads, each sending and receiving data via the http client API to communicate with a Web server (in the testing lab) to consume the network bandwidth. The variable MAX_NETWORK_THREADS is the number of network threads needed to consume 100% of the available network bandwidth. Then, it controls the network bandwidth consumption by starting a portion of such threads, i.e., MAX__USB_THREADS × percentage ÷ 100.

TAST also provides an API to determine a value for MAX_NETWORK_THREADS, which simply allocates an increasingly number of threads (initially 1) until all the available network bandwidth has just been consumed. This algorithm is useful for testing network-dependent applications such as online music players, online video streaming applications or Web browsers. Instead of freezing and buffering endlessly, these applications should either reduce their bandwidth requirement or stop gracefully under adverse network conditions.

The algorithm `consumeUSB()` controls the USB bandwidth. It is identical to `consumeNetwork()`, except the following. First, the variable MAX_NETWORK_THREADS is replaced by a new variable MAX__USB_THREADS, which is the number of reader/writer threads needed to fully saturate the USB bandwidth. Second, the algorithm starts a writer thread and a reader thread in turn. Due to the similarity between the two algorithms, for brevity, we skip to present `consumeUSB()` explicitly. Moreover, TAST has a method that determines a value for MAX__USB_THREADS, whose strategy is similar to the method that determines MAX_NETWORK_THREADS described above. TAST has a method that consumes a certain percentage of the USB storage space, which simply writes dummy contents to new files kept by the USB device.

In summary, to improve the automation support of MT, TAST has been extended to include 10 API methods to query and set up the amount of resources consumed in the execution environment.

### E. Events for Voice Control via Socket-based Voice Interface

In this section, we present the handling of events for voice commands. Test engineers may manually use the microphone on the remote controller of a TV to send a sound stream to the voice assistant application of the TV. TAST on the other hand converts the parameters (voice commands) of the voice-related methods in a test script into audio streams and send these streams via the interface to the voice assistant application.

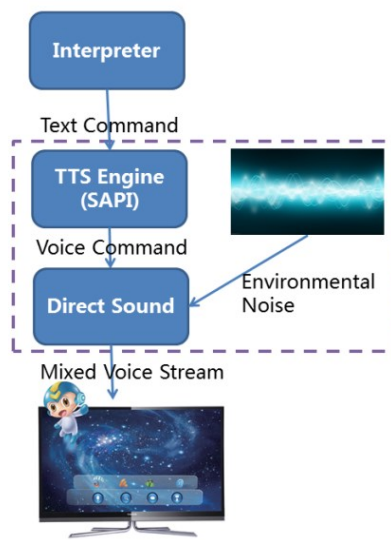

Figure 5. Workflow for methods sendVoiceCmd() and sendNoisyVoiceCmd() to help test applications with voice assistance.

As shown in Figure 5, if the voice controller receives an event for the MIC interface from the interpreter, it uses a Text-To-Speech (TTS) library to convert the text into a voice stream. As a requirement of Changhong, TAST is currently configured to use The Microsoft Speech API as the TTS library. Optionally, the voice controller then uses the DirectSound API for audio mixing so that a noisy home environment can be simulated. Finally, the resultant audio stream is sent to the voice assistant application via socket interface using a custom protocol. TAST provides methods `sendVoiceCmd()` and `sendNoisy-VoiceCmd()` for perfect and noisy environment, respectively.

## IV. CASE STUDY

In this section, we present the empirical study to answer the important research questions.

*A. Background*

The case study was conducted at the Reliability Institute of Changhong at Sichuan Province, China during Mar–Sep 2013. This institute is the testing arm of the company and is responsible to test various types of electronics products manufactured by the company, including smart TVs. The company has used the MT (MobileTest [7]) tool since 2008, which is an evolving platform that adds Android application testing support since 2010. Since the testers are quite familiar with the test development environment of MT, they ask us to continue improve MT to better support Android TV applications testing in this project.

The case study started with a series of requirement elicitation workshops between the researchers and more than 20 test engineers of the institute to identify the requirements of a testing tool. The company had developed a few smart TV models before this project commenced in 2013. Ten test engineers were eventually selected to join this project. Each was assigned with two applications to test. The list of applications (see Table II) for the smart TV model had been predetermined by the company.

As presented in Section I, we refer the original version of the testing tool used by the test engineers to as **MT**, and the final version of the same testing tool to as **TAST**.

For each such application, the test engineers had written in natural language a set of *planned test cases* (and sometimes, we refer to them as *test specifications* in this paper) for these existing smart TV models. Most of them were in the form of documentation for test engineers to interpret and follow them to conduct manual testing.

Despite the significant value of industrial case study used for evaluation [12], we should note here that an industrial case study did not enjoy the same level of rigor and freedom as a controlled experiment conducted in a research laboratory. For instance, the number of test engineers in the testing project, the training schedule of the TAST framework, the available time budget for test case design and execution, the choice of TV models and applications, and other factors must conform to the actual setting of the involved industrial projects. However, an industrial case study can truly testify to what extent a research proposal can handle real-world setting and identify limitations for future research.

*B. Research Questions*

We aim to study three refined research questions through the case study.

**RQ1:** To what extent can the planned test cases be completely developed on TAST as *fully automated* test scripts for the stress testing of Android-based applications of a targeted smart TV model?

**RQ2:** To what extent can TAST-automated test cases be completely reused when conducting stress testing on the same application but installed in another smart TV model of the same TV series?

**RQ3:** Can TAST-automated test cases effectively detect unknown faults from the targeted TV models?

*C. Application Benchmarks*

In the case study, the institute provided data of 20 Android-based TV applications listed in Table II for us to study the three research questions. To the best of our knowledge, these applications were also some of the most heavily tested applications by the testing team of the institute, and they collectively represented a set of most frequently used applications of a smart TV manufactured by the company.

*D. Experimental Setup*

The researchers *passively* monitored a real-world testing project on a real-world TV model `3D42A7000iC (L47)`. This model was one of the latest smart TVs running Android 4.0.1 manufactured by the company within the case study period. At the time of reporting the case study in this paper, this TV model is being sold in the retail market of China. We note that this is a real testing project that the test engineers must conduct testing according to the planned test cases (using their own approaches).

Ten (10) full-time test engineers were assigned to this testing project to conduct the system test. We ran a two-day training workshop session to educate these test engineers on how to operate the testing tool to write and run test scripts. They also learned the coding approach to data-driven testing (i.e., to keep data in a data source, and populate the parameters in a test script by retrieving data rows from the associated data sources). Our testing engineers were experienced on smart TV testing. They have evaluated various testing platforms including Robotium, APPium, and Sikuli before adopting the approach presented in this paper. Several of the testing engineers also had experiences on using Robotium for writing test scripts for three months in a previous testing project.

Specifically, each test engineer examined each planned test case of each application assigned. Based on his/her own professional experience, if the test engineer considers the planned test case can be automated by MT, he/she labeled the MT test script *automated*, otherwise, *manual*.

Next, the test engineer attempted to implement each planned test case as a TAST test script. As expected, some test cases were eventually still not fully supported. If a test engineer found the development of the corresponding test script was *either* not possible *or* too difficult to implement and decided not to automate it using TAST, the test engineer marked the test case as *manual*, otherwise *automated*. The test engineers then executed each test script thus produced on the given TV model.

According to their feedback after the case study, TAST is easy to use and the time cost of creating TAST-based test cases is small. They did not record the time spent on individual test cases, however. It is because in the industry setting, test engineers usually and flexibly switch between multiple tasks (e.g., answering unplanned calls or queries from others or attending meetings). Such situation made accurate recording the time spent in detail impractical.

To answer RQ1, we measured the ratio of "automated" test scripts to the total number of planned test cases with respect to each application. To ease our presentation, we refer to such an "automated" test script marked by TAST and MT as *TAST-automated* and *MT-automated* test scripts, respectively.

Table II Benchmarks and experimental results on planned test case automation

| App ID | Applications | Description | Planned (A) | # of test cases automated | | In ratio | | |
| --- | --- | --- | --- | --- | --- | --- | --- | --- |
| | | | | TAST (B) | MT (C) | D = B / A | E = C / A | F = D – E |
| 1 | TV Control 1.0.1 | TV control to set channel, volume, etc | 75 | 64 | 46 | 0.853 | 0.613 | 0.240 |
| 2 | Setting 1.2.1 | TV setting app | 82 | 79 | 49 | 0.963 | 0.598 | 0.366 |
| 3 | Browser 29.0.1547.23 | A built-in web browser for TV | 116 | 94 | 59 | 0.810 | 0.509 | 0.302 |
| 4 | Market 1.1.0 | A built-in app market for TV | 73 | 58 | 50 | 0.795 | 0.685 | 0.110 |
| 5 | Weather Report 1.2.3 | A built-in weather report app | 79 | 66 | 42 | 0.835 | 0.532 | 0.304 |
| 6 | Online Video 4.5.3 | Online video streaming app | 94 | 89 | 67 | 0.947 | 0.713 | 0.234 |
| 7 | Media Player 2.72 | A video/audio player that play movies on the USB storage | 88 | 79 | 47 | 0.898 | 0.534 | 0.364 |
| 8 | Photo Viewer 3.2.1 | An app to view photos | 53 | 49 | 32 | 0.925 | 0.604 | 0.321 |
| 9 | Temple Run 5.1.2 | Temple Run [35] is a popular "Endless Running" game | 79 | 74 | 43 | 0.937 | 0.544 | 0.392 |
| 10 | Ciri 1.2.5 | Ciri [28] is a voice assistant app developed by Changhong | 91 | 77 | 49 | 0.846 | 0.538 | 0.308 |
| 11 | Dictionary 2.7 | A Google Translate app | 73 | 66 | 45 | 0.904 | 0.616 | 0.288 |
| 12 | Calendar 201305280 | A calendar app | 61 | 52 | 38 | 0.852 | 0.623 | 0.230 |
| 13 | Weibo 3.6.0 | A Weibo client on TV | 76 | 67 | 51 | 0.882 | 0.671 | 0.211 |
| 14 | IM 1.3.5 | An instant messenger on TV | 41 | 39 | 23 | 0.951 | 0.561 | 0.390 |
| 15 | News 2.4.0 | A NetEase news client | 92 | 69 | 53 | 0.750 | 0.576 | 0.174 |
| 16 | Map 7.0.1 | A built-in map app | 89 | 70 | 45 | 0.787 | 0.506 | 0.281 |
| 17 | Calculator 1.7.3 | A built-in calculator app | 80 | 73 | 40 | 0.913 | 0.500 | 0.413 |
| 18 | File Explorer 2.3.0 | A built-in file explorer app | 43 | 39 | 24 | 0.907 | 0.558 | 0.349 |
| 19 | Car Race 5.1.0 | A car racing game app | 77 | 63 | 48 | 0.818 | 0.623 | 0.195 |
| 20 | Email 1.2.5 | An email client app | 101 | 80 | 65 | 0.792 | 0.644 | 0.149 |
| | | Total | 1563 | 1347 | 915 | – | – | – |
| | | Mean | 78 | 67 | 46 | 0.868 | 0.587 | 0.280 |
| | | Standard Deviation | | | | 0.060 | 0.059 | |

After they have completed the above testing sessions, the institute provided a new smart TV model for us to evaluate the reusability of the test scripts developed by the test engineers. This new TV model was equipped with a different hardware configuration and an upgraded Android OS version (Android 4.1). Note some of the applications also update accordingly to this new TV model. The test engineers ran each TAST-automated test script and each MT-automated test script on this new model.

To answer RQ2, we measured the ratio of such test scripts that each can successfully be run on this new TV model without any modification to the total number of "automated" test scripts.

Each TAST test script contained one or more assertion statements (e.g., line 25 in Figure 1) as test oracle. We also analyzed the dataset for RQ1 to identify whether there were any test scripts that violated any assertion statement or threw any exceptions in the course of execution. Each of such test scripts is marked as a *failed* test script.

To answer RQ3, we measured the number of failed test scripts to the total number of TAST-automated test scripts. The TV models are real products. As such, the development engineers had also debugged the applications (or else, the TV model would not be launched). We requested the development engineer to share with us some representative root causes (i.e., the faults) of the failures found that have not been detected without using TAST beforehand.

## V. DATA ANALYSIS

### A. Answering RQ1: Extent of Automation

The fourth to sixth columns in Table II show the number of planned test case, the number of TAST-automated test, and the number of MT-automated test scripts, respectively. We found that on average, TAST and MT successfully automated 87% and 59% of all the planned test cases. We also found that for each application examined, the number of TAST-automated test cases was more than the corresponding number of MT-automated test cases by 10% to 41%, with an average of 28%. This margin of difference is significant.

We have further performed an ANOVA test using MATLAB to confirm whether the two datasets are different significantly from each other at the 5% significance level. The ANOVA test yielded a *p*-value of $1.11 * 10^{-6}$ (with SS = 0.7952 (error = 0.1473), df = 1 (error = 38), MS = 0.7952 (error = 0.0039), F = 205.1396), which successfully rejected the null hypothesis that there was no difference between the two groups of data at the 5% significance level.

Although our methodology aims to automate code fragments that can be refactored as generic methods, this strategy may still be inadequate to make a test script fully automated. We thus conducted code inspection on the test scripts to understand why there were more TAST-automated test scripts than MT-automated test scripts.

On *Ciri*, TAST automated most of the planned test cases that were voice-driven, whereas MT had to leave such test cases for manual testing. This finding is a natural consequence of the API enrichment offered by TAST.

On *TV control*, *Setting*, *Weather Report*, *Dictionary*, *Calendar*, *News*, and *Email*, TAST automated significantly more test cases than MT because the executions of those test cases often led to small changes in only a few user-interface (UI) controls. Like other capture-and-replay tools, MT used an image comparison approach to checking test results. A small difference in the image would make MT to produce unreliable test reporting if the test engineers did not examine each difference and wrote specific comparison scripts to get rid of this false positive difference between the images under comparison. In a many cases, test engineers gave up automating such test cases due to the tedious code development and resolved to use their "eyeball" to scan the TV screens. TAST provided properties of UI controls to provide data for test oracle checks, which was more logical to be handled, more precise in item location comparison, and more insensitive to the changes in UI control irrelevant to the selected properties.

On *Market, Online Video, Media Player, Photo Viewer, Temple Run, Weibo, IM, Map, Calculator,* and *Car Race*, quite many planned test cases required resource controls for stress testing of the applications under diverse CPU, memory, and/or USB bandwidth conditions. Like other capture-and-replay tools, MT had no mechanism to automate the resource control function, and the test engineers chose to give up the automation of these test cases. (Note that in some test cases, the test engineers actually wrote some Linux shell scripts to ease their testing while using MT.) On *Browser* and *File Explorer*, TAST automated significantly more test cases than MT also due to the above reasons or their combinations.

We had also measured the numbers of lines of code for the TAST-automated test scripts and MT-automated test scripts. We found that the two types of test scripts resulted in 56 to 378 lines of code with a mean of 120 and a standard deviation of 30, and 80 to 460 lines of code with a mean of 223 and a standard deviation of 75, respectively. The result indicated that TAST-automated test cases tended to be significantly coarser in granularity to complete the same tasks. The result is consistent with the purpose of the testing tool enhancement to provide a higher-level test-script development platform than MT to assist test engineers.

Some test specifications were not automated. We found that almost all of them were due to their reliance on human involvement. The test engineers deemed them to be too hard to automate. Examples include human interactions with hardware, the verification of video quality, and the evaluation of user experience, which were either too costly to implement or subject to human judgment. For example, one test specification requires "*unplug the USB disk when playing movie stored in the USB, and plug the USB again to check whether the movie file can still be played (or is corrupted)*". This specification was not automated due to the "unplug and then plug" actions. To automate these test case specifications further, we may use a robot and further integrate TAST with it. However, it is beyond the scope of this work.

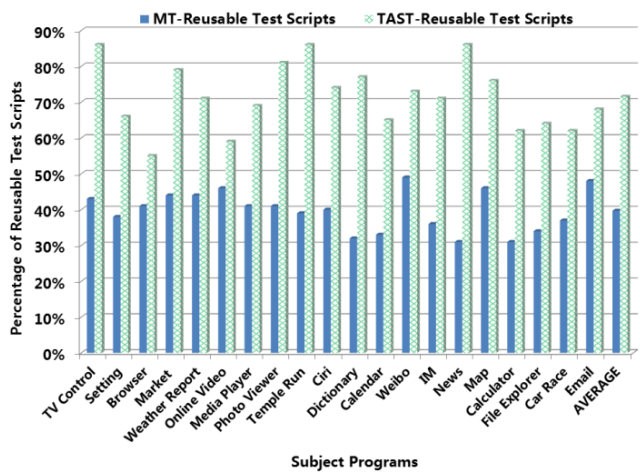

Figure 6. Comparison of the percentage of reusable test cases

Table III. Number of TAST Reusable Test Scripts Covering and Without Covering Changes

| Applications | TAST Reusable (A) | TAST Reusable Scripts Covering Change (B) | TAST Reusable Scripts Covering No Change (C) | In Ratio D= B/A | In Ratio E= C/A |
|---|---|---|---|---|---|
| TV Control | 55 | 20 | 35 | 37% | 63% |
| Setting | 52 | 20 | 32 | 38% | 62% |
| Browser | 51 | 20 | 31 | 40% | 60% |
| Market | 45 | 17 | 28 | 38% | 62% |
| Weather Report | 46 | 18 | 28 | 40% | 60% |
| Online Video | 52 | 19 | 33 | 37% | 63% |
| Media Player | 54 | 21 | 33 | 39% | 61% |
| Photo Viewer | 39 | 15 | 24 | 39% | 61% |
| Temple Run | 63 | 26 | 37 | 41% | 59% |
| Ciri | 56 | 21 | 35 | 37% | 63% |
| Dictionary | 51 | 19 | 32 | 38% | 62% |
| Calendar | 33 | 13 | 20 | 40% | 60% |
| Weibo | 48 | 18 | 30 | 37% | 63% |
| IM | 27 | 10 | 17 | 38% | 62% |
| News | 59 | 24 | 35 | 41% | 59% |
| Map | 53 | 19 | 34 | 37% | 63% |
| Calculator | 45 | 18 | 27 | 40% | 60% |
| File Explorer | 25 | 10 | 15 | 38% | 62% |
| Car Race | 38 | 15 | 23 | 40% | 60% |
| Email | 54 | 21 | 33 | 39% | 61% |
| Mean | 48 | 18 | 30 | 37% | 63% |
| Standard Deviation | 10 | 4 | 6 | 1% | 1% |

B. *Answering RQ2: Extent of Reuse*

Figure 6 shows the percentage of TAST-automated test scripts that were successfully reused in testing the same application on a newer TV model, and MT-automated test scripts alike. We refer to them as **TAST-Reusable** test scripts and **MT-Reusable** test scripts, respectively. The x-axis represents the benchmarks; and the y-axis is the percentage of test cases that have been successfully reused.

For each benchmark, there is a pair of bars. The solid bar in blue color on the left is for MT-Reusable test scripts, and the bar (in green) with a pattern filled on the right represents the TAST-Reusable test scripts. The rightmost pair of bars show that the mean percentage of reusable test cases over all benchmarks. The result shows that TAST resulted in 55.0–88.0% (with a mean of 71.4%) reusable test scripts, compared to 31.0–49.0% (with a mean of 40.7%) when using MT. On average, the difference is 30.7%, which is significant. However, the standard deviation achieved by TAST is 11.78%, which is higher than that of MT (5.36%).

Table IV. Sample Cases of Reusable Test Cases

| | MT | TAST | Example |
|---|---|---|---|
| Reuse | Failed | Failed | One test script is to test the TV system setting application of TV Model A. However, for TV model B, the TV system setting application was changed significantly, where both the Activity class name and the identities of controls changed, which made the test scripts of both tools invalid immediately. |
| | Successful | Successful | One test script is to handle the change in D-pad hardware due to the change in TV model. Although the keys were the same set of keys logically, yet the raw codes of some D-pad keys changed due to driver updates. In such scenario, the interpreter layers of both MT and TAST internally updated their implementation to use the valid raw codes when testing the new TV model. Because each of these two interpreters still provided the same set of testing APIs, the test scripts produced can still be reusable on both frameworks. Hence, the hardware changes had been successfully hidden by the layered design of both MT and TAST. We believe that the stability of the testing API provided by the interpreter layer across TV models was the key to this reusability. |
| | Failed | Successful | Due to the OS customized update, the ADB command waiting for device connection may fail to connect in WiFi environment occasionally. Since MT APIs are only low-level wrapper over ADB commands, the test script becomes non-reusable as it fails to handle the exception case in the test script. In contrast, the TAST scripts APIs are relatively high-level, which handles such potential ADB failure in the API implementation. As such, the high-level abstraction of the TAST testing APIs makes test scripts less vulnerable to underlying changes. |
| | Failed | Successful | One test script of this type is to allow changes in the positions of the controls of application GUI across TV models, but the identities of these controls remained unchanged. TAST test script only referenced the controls by identities, and so, the test scripts were still usable. However, the MT test script referenced the controls by their positions on screenshots, which made the MT test script non-reusable. We observe that in this scenario, the reusability of TAST over MT was due to the test engineers' knowledge on GUI information and the use of identity-based control reference in test scripts, where were supported by TAST but not in MT. |
| | Failed | Successful | Yet another case where TAST beats MT in terms of reusability is attributed to its test result checking strategy. MT uses image comparison for test results verification while TAST uses assertion statements over GUI control properties for the same purpose. One test script is simply to launch the weather application and then verify whether the default city is "Beijing". In the case study, the weather application was updated to use a new background. As a result, the MT test script became invalid because the image comparison approach always considered that the two images differ significantly. On the other hand, the TAST test script extracted the text label of the City control, and used a string comparison to check whether it is "Beijing". Therefore, the TAST script was immune to the change in application background. |
| | Successful | Failed | No such case observed in the case study. |

In last section, we observe that TAST is able to automate significantly more test cases than MT, and hence the above difference may be conservative. If we took the percentage of automated test cases into account, then on average, test engineers were able to completely reuse 61.9% (= 86.8% × 71.4%) and 23.9% (= 58.7% × 40.7%) of all test cases via TAST and MT, respectively. Encouragingly, the difference is 38%.

Moreover, for each benchmark, we also found that the bar for TAST-automated test script was significantly longer than that for MT-automated test scripts, indicating that the degree of reuses provided by TAST was usually significantly higher than that provided by MT.

Similar to what we did in the data analysis for RQ1, we also conducted an ANOVA test to confirm whether the amount of reuse between the two types of automated test cases is different significantly at the 5% significance level. The ANOVA test yielded a $p$-value of $1.04 * 10^{-5}$ (with SS = 1.0240 (with error = 0.2617), df = 1 (with error = 38), MS = 1.0240 (with error = 0.0069), and F = 148.6837), which successfully rejected the null hypothesis that there was no significant difference between the two groups of data.

The finding shows that the use of our APIs can not only improve the amount of automation (see RQ1) but also provide a higher potential of test case reuse. This finding has a strong implication to model-based program testing because its clear advantage is the ability to reuse test cases across different programs under test.

We found there are 5 possible sources of changes affecting script reusability in this case study. The change of hardware support (e.g., codec), the new constraints imposed by OS upgrade (e.g., local SD card file access), the change of application logic (e.g., the class name), the change of application UI (e.g., the position or layout of the controls), and a mixture of them may all affect the reusability of a test script.

Table V. Sample Failure Cases

| Failure Case | How to Repeat | Result | Discussion |
|---|---|---|---|
| 1 | Start a web browser. Go to youku.com. Actively consume 70% all available network bandwidth. Select to play a video. Press the "fast forward" button twice. | An ANR error occurs, showing the web browser not responding. | This is a typical error that the application using network cannot handle these limited bandwidth scenarios gracefully. With the network bandwidth control support of TAST, this bug can be exposed easily. |
| 2 | Actively consume 80% CPU usage. Start Temple Run. Start playing the game for a while. | The game renders very slowly, and gets stuck from time to time. Finally an ANR error occurs, showing Temple Run not responding. | Temple Run is a computation-intensive application. The error occurs only when the available CPU resource is stringent. With the CPU usage control support of TAST, it is easy to set up different levels of CPU utilization to expose this bug. |
| 3 | Actively consumes the USB storage space so that the available storage space is less than 1 Megabyte. Start Market, Select the application Angry Bird (which is larger than 10 Megabytes in size) to download from Market. Select to save Angry Bird in the USB disk. | Instead of showing a reminder dialog box expressing that there is not enough space to store Angry Bird, the Market application simply crashes. | When saving files to an external storage, an application should check whether an enough space is available. Market fails to perform this kind of check before writing data to the USB disk. Previously, test engineers have to prepare an almost full USB disk manually. With the support of TAST, the USB disk storage can be prepared with a single line of code in the test script. |
| 4 | Start Ciri. Read the names of the all applications in the TV from a database into a list. Iterate in turn over the list to instruct Ciri to launch the application by saying each name in the noisy mode. | Ciri cannot launch the correct application in the noisy mode. | The accuracy of the voice assistant application degrades seriously when the word is spoken in a noisy environment. Without the support of TAST, the test engineers have to set up an environment to prepare a noisy scenario. Furthermore, in the above steps, TAST can allow test engineers' to check whether the correct application has been activated with the support the Agent service (similar to line 25 in Figure 1) |
| 5 | Consume the memory until the system issues a low memory alert. Open the TV setting application. Repeatedly sets volume, picture room, signal source, and subtitles for 10 minutes. | The setting application crashes after 5 minutes. | There is a memory leak bug in the TV setting application. This bug only manifests when the application is under stress testing for a long period. Without controlling the memory usage, the setting application can run for 2 hours without crash. However, when the test engineers execute the application under a memory constrained scenario, the bug can manifest into a crash significantly quickly. |

We have further analyzed these TAST-reusable test cases to check whether they have covered any changes. As shown in Table III, the first column lists the applications and the second column lists the number of reusable test scripts by TAST. The third and the fourth column list the number of TAST reusable test scripts covering/without covering changes. The fifth and the sixth columns list the percentage of TAST reusable test scripts covering/without covering changes with respect to all TAST reusable test scripts, respectively. The last two rows list the mean and standard deviation of the corresponding column, respectively.

In general, we found that 59–63% of the reusable test scripts did not cover any change on the newer version of the TV model; and the remaining 37–41% of the reusable test scripts cover some changes each and still operated correctly. The mean number of TAST reusable test scripts over all applications is 48, of which 18 covering changes while 30 covering no changes, on average.

We had carefully examined the TAST scripts to understand why those TAST scripts survived the changes.

For *TV control*, *Setting*, *Ciri*, *Dictionary*, *Calendar*, *IM*, *Calculator*, and *File Explorer* applications, their scripts heavily interact with the underlying Agent interfaces. Fortunately, the high-level abstraction of the TAST testing APIs encapsulates the changes in the hardware and operating system well, which makes the test scripts less susceptible to those changes.

For *Browser*, *Market*, *Weibo*, *News*, and *Email* applications, they involve small UI changes in their relatively complex user interface, the position independent control id reference used in the TAST scripts makes them insensitive to small layout changes in GUI. In contrast, the MT scripts using position-based control reference are very susceptible to those small GUI changes.

For *Online Video*, *Media Player*, *Photo Viewer*, *Temple Run*, *Map*, *CarRace*, and *Weather Report* applications, they have relatively rich multimedia content. In MT, the testers are can only use image comparison for results verification. However, in TAST, they have the option to perform test results checking with assertions on GUI control properties. This gives TAST scripts great advantages over MT scripts in terms of reusability.

Table IV illustrates cases where at least one of TAST and MT successfully make the test script of the same test specification reusable. In particular, we did not find any test specification that MT made it reusable but TAST failed. In the table, the second and the third columns show whether MT and TAST successfully reused the same test specification, and the fourth column illustrates an example of the categories.

We also note that TAST uses the properties of widgets as the source of test oracle information. This strategy may not generally applicable, such as when the widget is incorrectly rendered on the screen or some other widgets shelter the target widget. To check faults like the inconsistency between a set of displayed widgets against their in-memory representations, the image comparison strategy may be more valuable. On the other hands, in our case study, we found that for many test specifications, TAST scripts did result in test cases with reusable implementation of test result checking while MT scripts did not. It may indicate that the use of the widget properties as the source of test oracle information can be one of the effective strategies.

To understand the situation better, we examined these TAST-reusable test scripts and classified them into 5 categories based on the sources of changes they covered. Specifically, we classified each of these test scripts based on the test script covered the change of hardware, OS, application logic, application UI, or a mixture of them. We list their ratios (in both mean and standard deviation) in Table VI. We observe from Table VI that these test scripts most frequently cover changes in application UI or logic, followed by changes in OS and hardware. Moreover, around 8% of these test scripts cover changes in more than one source (as indicated by the rightmost column of the table). Finally, the standard deviations of the ratios are consistently no more than 5%, which is small.

Table VI. Categorization of TAST-Reusable Test Scripts

| Category | hardware | OS | application logic | application UI | mixed |
|---|---|---|---|---|---|
| Mean | 12% | 16% | 28% | 36% | 8% |
| Stdev. | 3% | 3% | 5% | 5% | 4% |

We further note that there is a tiny fraction of test cases that are non-reusable but without covering any change. We found that they were likely due to some factors affecting application execution but not completely controllable while the case study was conducted. For example, in the case study, the application had to interact with external inputs such as Internet data and TV signal channels. The congestion of Internet connection and the interrupted TV signals may lead to unexpected outcomes that invalidates the corresponding test scripts.

In summary, based on our analysis on the TAST test scripts, we believe that our code refactoring approach, the high-level abstraction of the TAST testing APIs, the position independent GUI control id reference, and the oracle checking strategy of TAST enabled a higher degree of test script reuse observed in the case study.

## C. Answering RQ3: Fault Detection Ability

We computed the failure rate, which was defined as the percentage of all TAST-automated test scripts that each exposed a failure. Figure 7 shows the failure rate of the TAST-automated test scripts in the entire testing project of the TV model. Note that it is a real testing project, any fault exposed has been fixed, and the same test script had been re-run to confirm the fault fixed. Hence, we deemed a test script exposed a failure if the test script ever detected at least a failure once in the entire testing project.

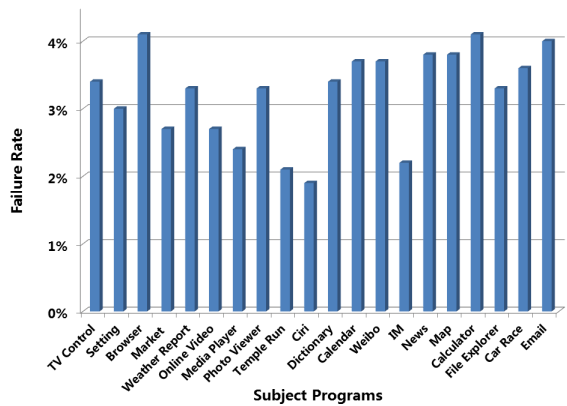

Figure 7. Failure rate of the developed test suites

We found that the TAST-automated test scripts were effective to expose faults in the given TV model. The failure rate achieved in the first round of testing ranges from 1.8% to 4.1% (with a mean of 3.2% and standard deviation of 0.68%), which was encouraging. It is because the planned test cases have been used to test these applications in earlier Smart TV models and all faults not so hard to be exposed have been cleared by the test engineers before our case study started. The result shows that TAST is likely to help the test engineers to improve the quality of the software component of the TV model. As a reference, Table V summarizes five selected failure scenarios detected by TAST when testing the given TV model.

## VI. RELATED WORK

There are many progresses toward full test automation and raise the level of abstraction to promote test script reusability. However, there is little research aiming to understanding how to raise the level of abstraction of test scripts. Indeed, to the best of our knowledge, our work is the first empirical study in the area of stress testing of TV applications.

Improving the test case reusability through repair is a popular strategy. Tiwari and Goel [20] reported a review study on reuse-oriented test approaches for reducing testing effort. Memon [14] proposed a GUI regression testing technique that performs automated test script reusability analysis on existing GUI test suites and fixing the non-usable test cases. *JAutomate* [1] used a screen-based image recognition approach to identify input locations and was able to simulate GUI control events (e.g., mouse movement) so that the test scripts can be reused to test applications in which GUI control objects have been relocated. Our study examines test case reusability from a model-based perspective. Our aim is to abstract test code so that at a certain abstraction level, there is no need to change the test code and the specific way to deal with a particular application under test is handled by the runtime engine of the testing tool internally.

In test automation, formulating effective test oracles must be addressed. Many existing GUI testing tools use image comparison for test result verification. Katona et al. [9] proposed an automatic and black-box technique to test TVs on a final product line. The technique used a camera to capture the image shown on a TV screen for test verification. Marijan et al. [12] proposed an automatic system for functional failure detection on TV. The system adopts an effective image comparison algorithm to compare the image captured from the TV under test and that from the reference TV. Their approach was applicable to test TV systems with or without operating system support. MobileTest [7] was a typical capture-and-replay testing tool based on image comparison. It used the ADB interface to stimulate the device under test and adopted the GUI screenshots for test result checking. Zhang et al. [23] compared two techniques addressing the test oracle problem: metamorphic testing and assertion checking. TAST provides test engineers with assertion checking APIs and can compare images and GUI control objects. We will study metamorphic testing for Android apps in the future.

A branch of test automation that is related to our study is the generation of test cases. Merkel [15] analyzed and enhanced the adaptive random test case generation technique, which is applicable to testing general applications. Pomeranz [16] proposed to build effective functional test sequences by concatenating test subsequences from a generated test pool. Jeon et al. [6] proposed a symbolic execution framework for the Dalvik bytecode. In this way, automatic test case generation through symbolic execution of Android byte code can be possible. Azim and Neamtiu [2] proposed to apply a static and taint-style dataflow analysis on the application bytecode to construct a high-level control flow graph that captured legal transitions among activities (app screens). They then performed systematic depth-first exploration on those graphs to generate events for testing Android applications. Choi et al. [3] proposed to uses machine learning to learn a model of the application during testing, uses the learned model to generate user inputs that visit unexplored states of the app. Jensen et al. [5] proposed a two-phase technique for automatically generating event sequences. They firstly used concolic execution to build summaries of event handlers of the application. Then, they used the summaries and the GUI model to generate targeted event sequences. Takala et al. [19] proposed a model-based approach for GUI testing of Android applications. Yeh et al. [22] proposed an approach to analyze GUI model during testing process, and then performed black-box Android testing based on the model.

Yang et al. [21] proposed a grey-box approach to automatic GUI-model generation of mobile applications. They performed static analysis to identify a set of events supported by the applications. Then, they systematically exercised the identified events on the application. As we have validated in our case study, test scripts without the consideration of resources often lead to significantly lower degree of automation. We believe that our work has a positive impact on the research on test case generation.

Apart from the resource constraint issues, our study also shows the feasibility that by a process of code refactoring, one can develop a testing approach that provides a significantly higher degree of test automation than without such a process. This approach is orthogonal to the test automation strategies presented in the above work, which make our work unique and complements to them.

There are several works on testing embedded systems. Koong et al. [10] presented an automatic white-box testing environment for multi-core embedded systems. Their system can perform unit testing, coverage testing, and multi-core performance testing based on source code instrumentation and code generation techniques. TAST is a black-box testing environment. Mattiello-Francisco et al. [13] proposed a technique for integration testing of the timing constraints of real-time embedded software. They used formal models to describe the timing and interoperability specifications for test case generation. Different from their work, ours study focuses on system level testing instead of integration testing. Satoh [17] proposed to test context-sensitive networked applications via emulating the physical mobility by the logical mobility of the underlying computing devices of these applications. Our strategy is to emulate the resource condition in the execution environment of the application in TV.

There are also several open-source or commercial testing tools for consumer electronics device testing. TestQuest [37] is a non-intrusive automated test solution that provides comprehensive support for a wide range of electronic devices. TestQuest executes predefined actions and compares the output to valid states to determine whether the test was successful by simulating a "virtual" user. The WindRiver UX Test Development Kit [39] is a test development environment targeted at GUI–based testing for Android platform. Wind River UX Test Development Kit is designed to assist in the validation of the user experience of a device by reproducing human interactions to test user interfaces.

Robotium [32] is a well-designed testing framework suited for both white-box and black-box testing of android application. However, several problems with it prevent us from selecting it. First, it uses the Android instrumentation test framework, which limits its execution within the same process of the application under test, so it can only work with activities and views within the defined package. This further makes the future extension of our tool to test the interaction of several applications (apks) within the same test case difficult. Second, the apk re-signing process required by Robotium for black-box testing is tedious. We adopt Monkey because it provides a clean and adequate interface to support our basic testing requirements.

The *monkeyrunner* [30] is a standard built-in tool within the Android SDK that provides APIs for writing programs that control an Android device or emulator from outside of Android code. It also uses Python as the scripting language. Owing to its immerse impact on the Android development community, TAST also includes several libraries of this tool as its build blocks (e.g., device connection management).

Testdroid [36] is a fully automated cloud suite for compatibility testing, facilitating Android developers to test applications on multiple real devices at the same time. It provides a Testdroid recorder tool for recording user actions, generating reusable Android JUnit test cases and running them in the Testdroid Cloud. The Testdroid Cloud provides an online service for testing applications on real Android devices at the backend.

The *Appium* [26] testing framework is an open-source cross-platform tool for automating native, mobile web, and hybrid applications on iOS and Android platforms. It is fair language-independent. But, its underlying implementation has constraints for Android app testing. On Android 2.3+, *Appium* depends on the Android's instrumentation framework, where the limitation on testing multiple applications still applies. On Android 4.2+, *Appium* depends on the *UiAutomator* [38] framework, which requires the applications under test to be designed with accessibility in mind [25]. However, not all third-party applications used in our case study provide accessibility support for their customized UI components. This limits the applicability of *Appium* in our testing scenario.

The *Sikuli* [33] framework automates GUI testing via the image recognition capability powered by *OpenCV* to identify and control GUI components. The writing of test scripts using *Sikuli* is simple and intuitive. But, in terms our own testing requirements, it also has some limitations. First, since image recognition is used, the reusability of the test scripts may be affected by small GUI changes (e.g., change of button style). Second, since it has no knowledge of system internal information, it cannot help control the resources available to an application for stress testing.

Reliability testing is a key concern in testing consumer electronic products. Marijan et al. [11] proposed to derive a model based on usage profiles on such products. They then performed a reliability analysis based on the execution results of the test cases derived from such a model. Different from this work, TAST mainly provides a script-based and black-box test infrastructure to develop and execute test cases. Huang and Lin [4] proposed to improve the software reliability modeling by taking the testing compressing factor and the failure-to-fault relationship into the model. Their evaluation on real failure data shows the proposed model having a good failure prediction power.

## VII. CONCLUSION

Many model-based testing strategies face the difficulty of translating their abstract test cases into concrete test cases. In this paper, we have examined this aspect from a reverse-engineering perspective. Specifically, we have reported an exploratory study. In the study, our testing methodology consisted of a number of steps to be conducted iteratively. First, we started from observing how test engineers wrote concrete test cases using an existing testing tool. Then, we identified pieces of code to become methods in the sense of "extract methods" in code refactoring, and identified missing features in the testing tool to prevent test engineers from automating the test scripts further. The testing tool was then enriched with such methods as APIs (with necessary runtime supports). This exploratory process continued until test engineers were (largely) satisfied with the current testing tools. We chose stress testing to study because fully automated test scripts were necessary, or else meaningful stress testing could not be conducted. Moreover, as newly automated test scripts can be applied to conduct a session of stress test on a program, owing to the combinatorial effect, it creates many new stress testing scenarios to test the program (together with existing automated test scripts). This testing methodology have been demonstrated in our case study that it exposed previously unknown bugs from the applications, thereby improving the reliability of the TV product, by producing an effective testing framework. Specifically, the case study has significantly demonstrated that at least 28% more test cases can be automated, 38% more test cases can be reused across TV models of the same TV series, and 3.2% of the automated test cases can expose previously unknown bugs in stress testing of the TV applications.

Our work also demonstrates a new research direction that bridging the gap between abstract test cases and concrete test cases can be a bottom-up process, which can be very effective. We have obtained many abstract test cases for each application. In the future, we will study how to formulate effective test models so that these abstract test cases can be automatically generated from such test models. In this way, we are one step closer to the goal of model-based program testing, where concrete and executable test cases can be effectively generated from test models directly.

## VIII. ACKNOWLEDGMENT

This research is supported in part by the National Natural Science Foundation of China (project no. 61202077), Civil Aviation Special Fund (project no. MJ-S-2012-05 and MJ-Y-2012-07) and the ECS and GRF of Research Grants Council of Hong Kong (project nos. 123512, 125113, 111313, and 11201114).

**Bo Jiang** is an assistant professor at School of Computer Science and Engineering, Beihang University. He got his PhD from The University of Hong Kong. His current research interests are software engineering in general, and embedded systems testing as well as program debugging in particular. He received the best paper awards from COMP-SAC'08, COMPSAC'09, and QSIC'11. His research results have been reported in many international journals and conferences such as TSC, JSS, IST, ASE, WWW, ICWS, COMPSAC, and QSIC. He is a member of IEEE.

**Peng Chen** is a MSc student at School of Computer Science and Engineering, Beihang University. His research interest is software quality assurance. He is a student member of the IEEE.

**W. K. Chan** is an associate professor at Department of Computer Science, City University of Hong Kong. He obtained all his BEng, MPhil and PhD degrees from The University of Hong Kong. He is currently editors of *Journal of Systems and Software* and *International Journal of Creative Computi*ng, guest editors of several international journals, PC/RC members of FSE'14 and ICSE'15, a symposium chair of COMPSAC, and an area chair of ICWS'15. His research results have been reported in more than 100 papers published in major international journals and conferences. His current research interest is to address the software testing and program analysis challenges faced by developing and operating large-scale applications. He is a member of IEEE.

**Xinchao Zhang** is a Senior Test Manager at Reliability Institute of Si-chuan Changhong Electric Co. in China. His research interest includes quality assurance and system reliability testing.